\documentclass[twoside]{article}

\usepackage{graphicx,fleqn,espcrc2}

\title{Transverse Spin Structure of the Nucleon from COMPASS}

\author{C. Schill on behalf of the COMPASS collaboration
\vspace{.3cm}\\
Physikalisches Institut der
Albert-Ludwigs-Universit\"at Freiburg\\
Hermann-Herder Str. 3, 79104 Freiburg, Germany.
}
       
\begin{document}

\begin{abstract}
The measurement of transverse spin effects in semi-inclusive deep-inelastic
scattering is an important part of the COMPASS physics program. In the years
2002-2004 data were taken by scattering a $160$~GeV muon beam on a
transversely polarized deuteron target. In 2007, additional data were collected
on a transversely polarized proton target. New results for the Collins
and Sivers asymmetries from the analysis of the proton data are presented.
\end{abstract}

\maketitle

\section{Introduction}
Single spin asymmetries in semi-inclusive deep-inelastic
scattering (SIDIS) off transversely polarized nucleon targets have been under intense
experimental investigation over the past few years \cite{COMPASS,HERMES,Bacchetta}. 
They provide new insights
into QCD and the nucleon structure. For instance, they allow the determination
of the third yet-unknown leading-twist quark distribution function $\Delta
_Tq(x)$, the transversity distribution \cite{Collins,Artru}. Additionally, they give insight into
the parton transverse momentum distribution and angular momentum
\cite{Jaffe}.

\section{The COMPASS experiment}

COMPASS is a fixed target experiment at the CERN
SPS accelerator with a wide physics program focused on the nucleon spin structure
and on hadron spectroscopy. COMPASS investigates transversity and the
transverse momentum structure of the nucleon in semi-inclusive
deep-inelastic scattering. A $160$~GeV muon beam is
scattered on a transversely polarized hydrogen or deuterium
target. The scattered muon and the produced hadrons are detected in a
wide-acceptance two-stage spectrometer with excellent particle
identification capabilities \cite{Experiment}. 
In the years 2002, 2003, and 2004, data were collected 
on a transversely polarized $^6LiD$ target. In the run $2007$  
data were taken with a transversely polarized $NH_3$ target. In the following
I will focus on the new results from the data taken on the proton.

\section{The Collins asymmetry}

In semi-inclusive deep-inelastic scattering the transversity
distribution $\Delta_Tq(x)$ can be measured in combination with the
chiral odd Collins fragmentation function
$\Delta_0^TD_q^h(x)$. According to Collins, the
fragmentation of a transversely polarized quark into an unpolarized
hadron generates an azimuthal modulation of the hadron distribution 
with respect to the lepton scattering plane \cite{Collins}. The hadron
yield $N(\Phi_C)$ can be written as:
\begin{equation}
N(\Phi_C)=N_0\cdot (1+f\cdot P_t\cdot D_{NN}\cdot A_C\cdot \sin \Phi_C),
\label{equ:Collins}
\end{equation}
where $N_0$ is the average hadron yield, $f$ the fraction of
polarized material in the target, $P_t$ the target polarization, and
$D_{NN}=(1-y)(1-y+y^2/2)$ the depolarization factor. The angle $\Phi_C$ is the
Collins angle. It is defined as  $\Phi_C=\Phi_h+\Phi_s-\pi$, the sum of the
hadron azimuthal angle $\Phi_h$ and the target spin vector azimuthal angle
$\Phi_s$, both with respect to the lepton scattering plane \cite{Artru2}. The
measured Collins asymmetry $A_C$ can be factorized into a convolution of the
transversity distribution $\Delta_Tq(x)$ of quarks of flavor $q$ and the
Collins fragmentation function $\Delta_0^TD_q^h(x)$, summed over all quark
flavors $q$:
\begin{equation}
A_C=\frac{\Sigma_q e_q^2\Delta _Tq(x)\Delta_T^0D_q^h(z, p_T^h)}
{\Sigma_q e_q^2 q(x) D_q^h(z, p_T^h)}.
\end{equation}
Here, $e_q$ is the quark charge, $D^h_q(z, p_T^h)$ the unpolarized fragmentation
function, $z=E_h/(E_\mu-E_{\mu'})$ the fraction of available energy carried by
the hadron and $p_T^h$ the hadron transverse momentum with respect to the
virtual photon direction. As can be seen from equation~\ref{equ:Collins}, the
Collins asymmetry shows as a $\sin\Phi_C$ modulation in the number of produced
hadrons.

\section{The Sivers asymmetry} 

Another source of azimuthal asymmetry is related to the Sivers effect. The
Sivers asymmetry rises from a coupling of the intrinsic transverse
momentum $\overrightarrow{k}_T$ of unpolarized quarks with the spin of a
transversely polarized nucleon \cite{Sivers}. The correlation between the transverse nucleon
spin and the transverse quark momentum is described by the Sivers distribution
function  $\Delta_0^Tq(x, \overrightarrow{k}_T)$. The Sivers effect leads to an
azimuthal modulation of the produced hadrons
\begin{equation}
N(\Phi_S)=N_0\cdot (1+f\cdot P_t\cdot A_S\cdot \sin \Phi_S).
\label{equ:Sivers}
\end{equation}
The Sivers angle is defined as $\Phi_S=\Phi_h-\Phi_s$. The measured Sivers
asymmetry $A_S$ can be factorized into a product of the Sivers distribution function and the
unpolarized fragmentation function $D_q^h(z)$:
\begin{equation}
A_S=\frac{\Sigma_q e_q^2\Delta_0^Tq(x, \overrightarrow{k}_T) D_q^h(z)}
{\Sigma_q e_q^2 q(x) D_q^h(z)}.
\end{equation}
In this case the asymmetry $A_S$ shows up as the amplitude of a $\sin\Phi_S$ modulation in the
number of produced hadrons. 

Since the Collins and Sivers asymmetries are independent azimuthal modulations
of the cross section for semi-inclusive deep-inelastic scattering \cite{Artru2,Boer}, both
asymmetries can be determined experimentally from the same dataset. 

\section{Data sample and event selection}
In 2007 COMPASS took data with a transversely polarized proton ($NH_3$) target.
The target consists of three cells (upstream, central and downstream) of 30, 60
and 30 cm length, respectively. The upstream and downstream cell are polarized
in one direction while the middle cell is polarized oppositely. The target
material has a high polarization of about $90$\% and a dilution factor of
$0.15$. The direction of the target polarization was reversed every five days.
The asymmetries are analyzed using at the same time data from two time periods
with opposite polarization and  from the different target cells.
For the results presented here, about $20$\% of the data collected have been
analyzed. The data have been selected requiring a good stability of the 
spectrometer within one and between consecutive periods of data taking. 

To select DIS events, kinematic cuts of the squared four
momentum transfer $Q^2>1$~(GeV/c)$^2$, the hadronic invariant mass
$W>5$~GeV/c$^2$ and the fractional energy transfer of the muon $0.1<y<0.9$ were
applied. The hadron sample on which the asymmetries are computed consists of all
charged hadrons originating from the reaction vertex with $p_T^h> 0.1$~GeV/c
and $z>0.2$. 

The Collins and Sivers asymmetries were evaluated as a function of $x$, $z$, and
$p_T^h$ integrating over the other two variables. The method used for the
extraction is based on a two-dimensional binning in $\Phi_h$ and $\Phi_s$. The
extraction of the amplitudes is then performed fitting the expression for the
transverse polarization dependent part of the semi-inclusive DIS cross section \cite{Boer}
to the measured count rates in the target cells by a maximum likelihood method,
taking into account the spectrometer acceptance. The results have been checked
by several other methods described in Ref.~\cite{COMPASS}.

A number of systematic studies have been performed to check against systematic
effects. The systematic errors have been estimated to be $0.3\cdot
\sigma_{stat}$ for the Collins and $0.5\cdot \sigma_{stat}$ for the
Sivers asymmetries.

\section{Results}

\begin{figure*}
\vspace*{-0.5cm}
\centerline{
\includegraphics[width=0.95\textwidth]{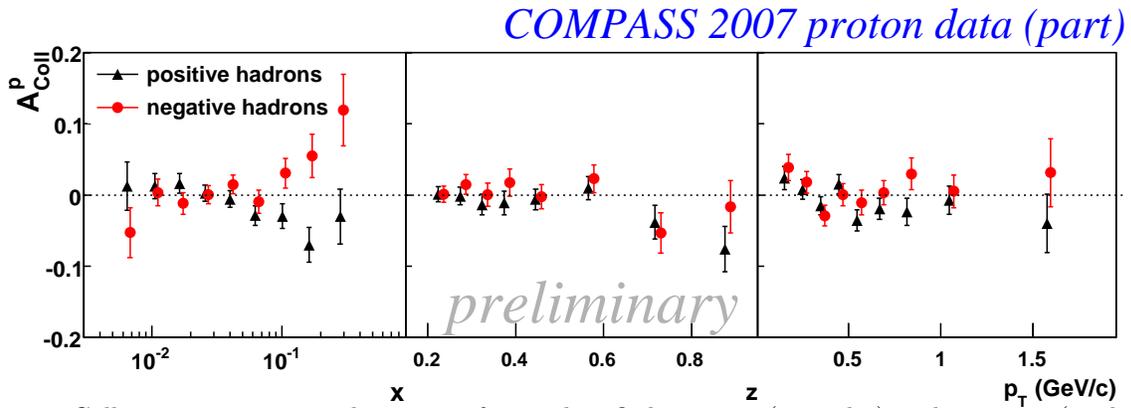}}
\vspace*{-1.5cm}
\caption{Collins asymmetry on the proton for unidentified positive (triangles) and negative
(circles) hadrons as a function of $x$, $z$, and $p_T^h$.}
\end{figure*}

\begin{figure*}
\vspace*{-0.5cm}
\centerline{
\includegraphics[width=0.95\textwidth]{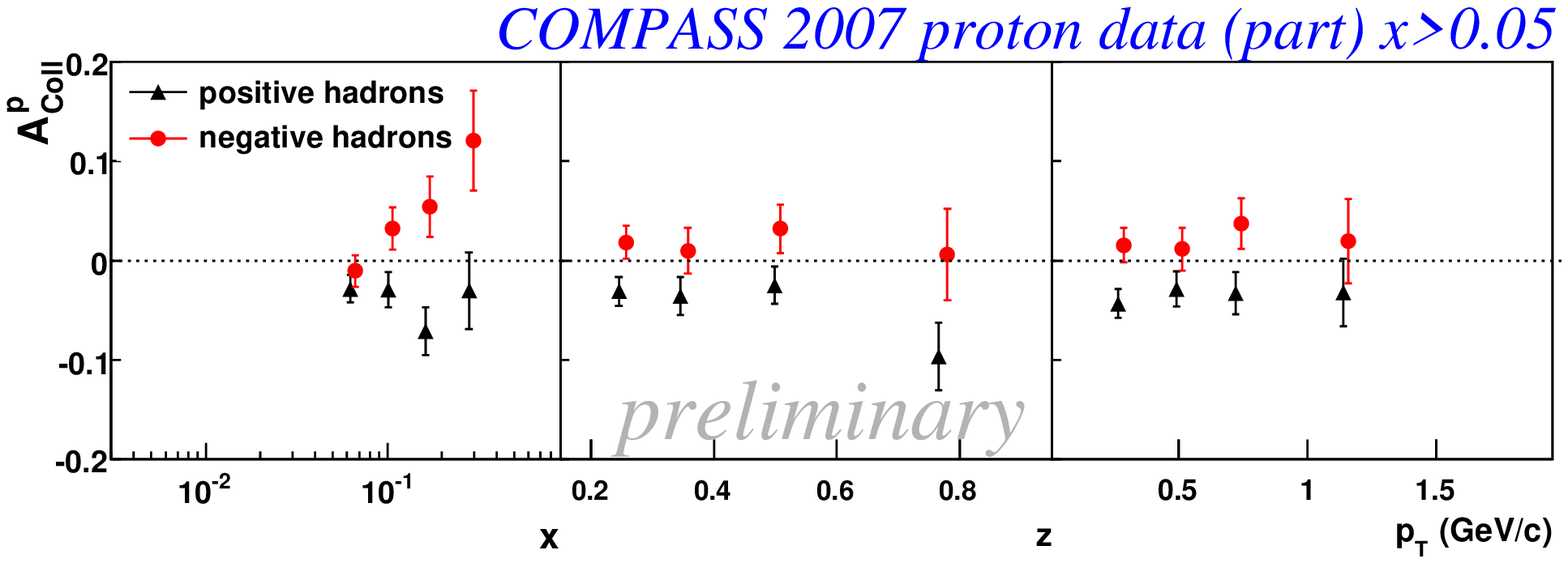}}
\vspace*{-1.5cm}
\caption{Collins asymmetry with the cut $x>0.05$  for unidentified positive (triangles) and negative
(circles) hadrons as a function of $x$, $z$, and $p_T^h$.}
\end{figure*}

\begin{figure*}
\vspace*{-0.5cm}
\centerline{
\includegraphics[width=0.95\textwidth]{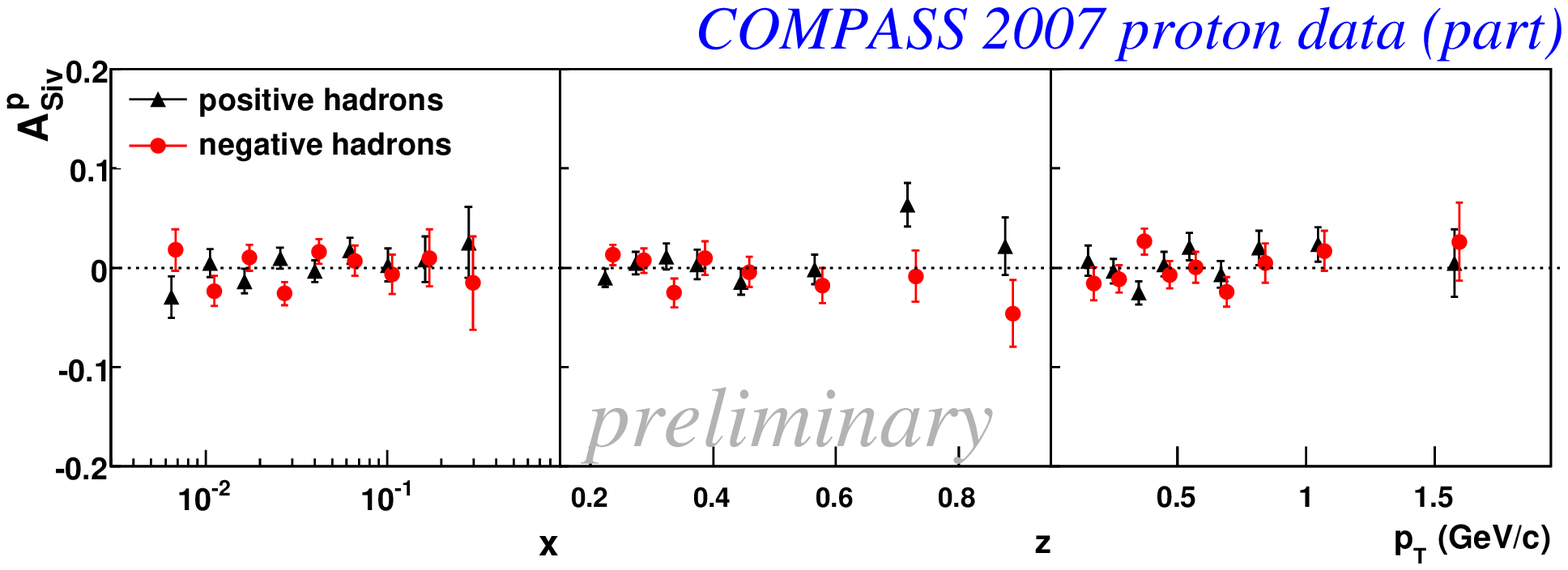}}
\vspace*{-1.5cm}
\caption{Sivers asymmetry on the proton for unidentified positive (triangles) and negative
(circles) hadrons as a function of $x$, $z$, and $p_T^h$.}
\end{figure*}

\begin{figure*}
\vspace*{-0.5cm}
\centerline{
\includegraphics[width=\textwidth]{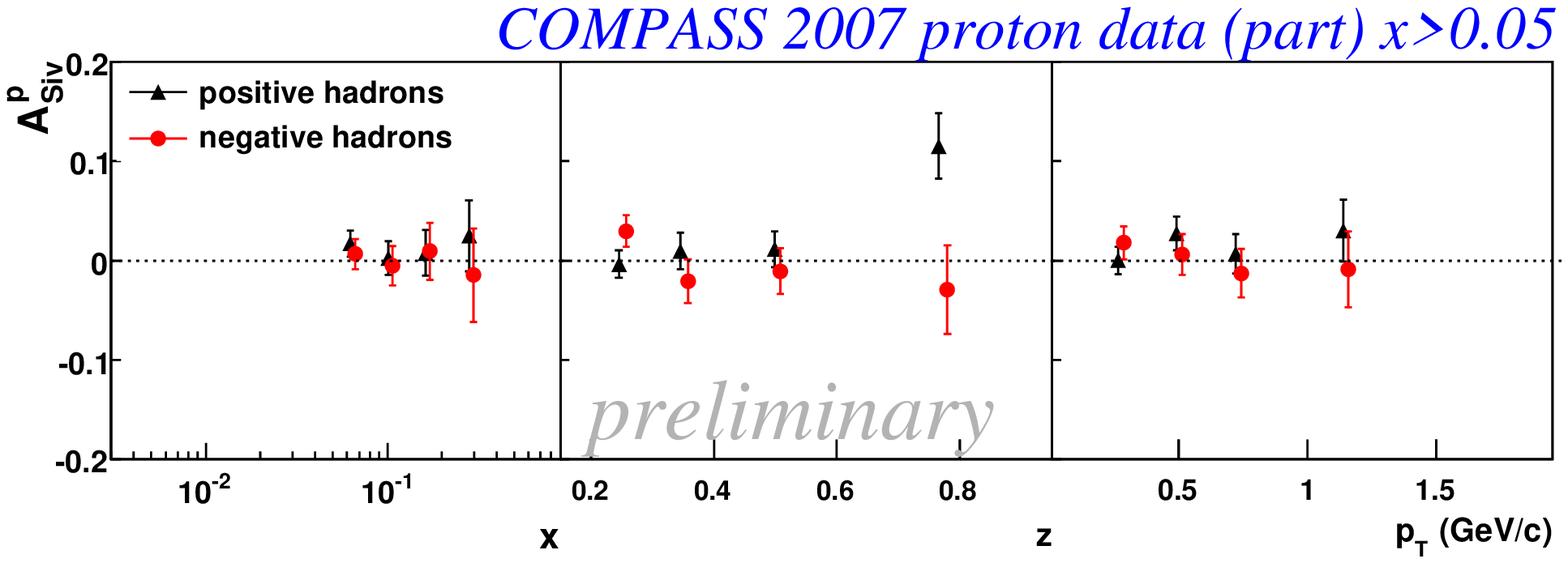}}
\vspace*{-1.5cm}
\caption{Sivers asymmetry with the cut $x>0.05$  for unidentified positive (triangles) and negative
(circles) hadrons as a function of $x$, $z$, and $p_T^h$.}
\end{figure*}

In Fig.~1 the results for the Collins asymmetry on
a proton target are shown as a function of $x$, $z$, and $p_T$ for positive and
negative unidentified hadrons. For small $x$ up to $x=0.05$ the measured
asymmetry is small and statistically compatible with zero, while in the last
points a signal is visible. The asymmetry increases up to about $10$\% with
opposite sign for negative and positive hadrons. Binned in $z$ and $p_T$  the
asymmetry is compatible with zero, since most of the statistical sample
is in the low $x$ region. Requiring $x>0.05$, the asymmetry signal becomes more
evident both in $z$ and in $p_T^h$ as shown in Fig.~2. There is
no appreciable $z$ or $p_T^h$ dependence.

In Fig.~3 the results for the Sivers asymmetry are shown as a
function of $x$, $z$, and $p_T^h$.  The Sivers asymmetry is small and
statistically compatible with zero for both positive and negative hadrons in 
all the measured $x$ range. The result for positive hadrons is in contrast to 
the measurement done by the HERMES collaboration \cite{Diefentaler}, and the two
data samples are only marginally compatible. Cutting at $x>0.05$ the measured
Sivers asymmetry on the proton is still compatible with zero within its statistical
errors, as shown in Fig.~4.

\section{Summary} 

New preliminary results for Collins and Sivers asymmetries measured at
COMPASS in semi-inclusive DIS on a transversely polarized proton target have
been presented. For $x>0.05$, a Collins asymmetry different from zero and with
opposite sign for positive and negative hadrons has been observed. Within
statistical precision of the measurement, the Sivers asymmetry is compatible
with zero, both for negative and positive hadrons. The dataset analyzed
corresponds to about $20$\% of the data recorded in 2007.

\end{document}